\documentclass[12pt,twoside]{article}

\usepackage{amsmath,amssymb,amsthm}

\newcommand{\Prim}{\mathrm{Prim}}

\newcommand{\Hyp}{\mathrm{Hyp}}

\newcommand{\tinf}{\to\infty}
\newcommand{\disp}{\displaystyle}
\newcommand{\bsla}{\backslash}
\newcommand{\nt}{\notag}

\newcommand{\bC}{\mathbb{C}}
\newcommand{\bR}{\mathbb{R}}
\newcommand{\bZ}{\mathbb{Z}}
\newcommand{\cS}{\mathcal{S}}
\renewcommand{\Re}{\mathrm{Re}}
\renewcommand{\Im}{\mathrm{Im}}

\DeclareMathOperator*{\Res}{\mathrm{Res}}

\newtheorem{thm}{Theorem}[section]

\newtheorem{lem}[thm]{Lemma}
\newtheorem{cor}[thm]{Corollary}

\newtheorem{definition}{Definition}[section]

\numberwithin{equation}{section}

\title{Hierarchy of the Selberg zeta functions}
\author{Yasufumi Hashimoto and Masato Wakayama}
\date{}

\begin{document}

\maketitle
\begin{abstract}
We introduce a Selberg type zeta function of two variables which interpolates several higher Selberg zeta functions.
The analytic continuation, the functional equation and the determinant expression 
of this function via the Laplacian on a Riemann surface are obtained.
\end{abstract}
\renewcommand{\thefootnote}{}

\section{Introduction}
Let $H$ be the upper half plane and 
$\Gamma$ a discrete co-compact torsion free  subgroup of $SL_2(\bR)$. 
The Selberg zeta function of $\Gamma$ is defined by
\begin{align}
Z_{\Gamma}(s):=\prod_{p\in\Prim(\Gamma)}\prod_{n=0}^{\infty}\big(1-N(p)^{-s-n}\big)\quad\Re{s}>1,
\end{align}
where $\Prim(\Gamma)$ is the set of the primitive hyperbolic conjugacy classes of $\Gamma$ 
and $N(p)$ is a square of the larger eigenvalue of $p\in\Prim(\Gamma)$. 
It is well known that  
the Selberg zeta function is analytically continued to the whole complex plane $\bC$ as an entire function 
and has the determinant expression (see, e.g. \cite{D'HP} and \cite{Sa}).
\begin{align}
Z_{\Gamma}(s)G_{\Gamma}(s)&=\det{\big(\Delta+s(1-s)\big)},\label{detexp}
\end{align}
where $G_{\Gamma}(s)$ is the gamma factor given as the product of Barnes' double Gamma functions, 
$\Delta=-y^2(\partial^2/\partial x^2+\partial^2/\partial y^2)$ is the Laplacian on $\Gamma\bsla H$ 
and the right hand side of \eqref{detexp} is the zeta regularized determinant of $\Delta+s(1-s)$. 
The functional equation $Z_{\Gamma}(s)G_{\Gamma}(s)=Z_{\Gamma}(1-s)G_{\Gamma}(1-s)$ 
follows immediately from this expression. 
Also, it is important to note that the analogue of the Riemann Hypothesis follows from the determinant expression,
because $\Delta$ is positive definite.

The function  
\begin{align}
\zeta_{\Gamma}(s):=\prod_{p\in\Prim(\Gamma)}(1-N(p)^{-s})^{-1}\quad\Re{s}>1
\end{align}
can be regarded as a Ruelle type zeta function and satisfies 
\begin{align}
&\zeta_{\Gamma}(s)=Z_{\Gamma}(s+1)/Z_{\Gamma}(s),\label{ruelle}\\
&Z_{\Gamma}(s)=\prod_{n=0}^{\infty}\zeta_{\Gamma}(s+n)^{-1}.\nt
\end{align}
Using \eqref{ruelle}, 
we easily see that $\zeta_{\Gamma}(s)$ is analytically continued to the whole complex plane $\bC$ 
as a meromorphic function and has the functional equation 
$\zeta_{\Gamma}(s)\zeta_{\Gamma}(-s)=(2\sin{\pi s})^{4(1-g)}$, 
where $g$ denotes the genus of the Riemann surface $\Gamma\bsla H$ (see, e.g. \cite{KW2}).

Moreover, the higher Selberg zeta function 
$z_{\Gamma}(s):=\prod_{n=1}^{\infty}Z_{\Gamma}(s+n)^{-1}$ was introduced in \cite{KW2} 
(see \cite{Mo} when $\Gamma$ is a congruence subgroup of $SL_2(\bZ)$). 
It was shown that 
$z_{\Gamma}(s)$ is meromorphically continued to the whole complex plane $\bC$  
and has the functional equation (involving the determination of the gamma factor).

In \cite{KW3} (see also \cite{Wi}), an analogue of the 
Casimir effect for 
a Riemann surface $\Gamma\backslash H$ was studied. 
Actually, the Casimir energy is explicitly evaluated by a definite 
integral of the logarithm of the Selberg zeta function 
$Z_\Gamma(s)$ with respect to $s$ (see (3.11) or (3.12) in \cite{KW3}). 
Then, a natural question how one can define the corresponding force 
arose. It is reasonable that one may obtain such force by taking a 
variation relative to $\Gamma\backslash H$ in the Teihm\"{u}ller space 
(the negative derivative of the energy gives the force!) 
In \cite{G1}, Gon shows that the first variation of the 
Selberg zeta functions is explicitly 
described by the (local) higher Selberg 
zeta function and a period of 
automorphic forms.

In order to establish a unified treatment of these zeta functions, 
we introduce a zeta function of two variables as 
\begin{align}
Z_{\Gamma}(s,t):=\prod_{p\in\Prim(\Gamma)}\prod_{n=0}^{\infty}\big(1-N(p)^{-s-n}\big)
^{\binom{t+n-1}{n}}\quad\Re{s}>1, t\in\bC,
\end{align}
which interpolates the higher Selberg zeta function of rank $r$ defined by
\begin{align*}
Z_{\Gamma}^{(r)}(s):=\prod_{n_1,\cdots,n_{r-1}=0}^{\infty}Z_{\Gamma}(s+n_1+\cdots+n_{r-1}).
\end{align*}
In fact, $Z_{\Gamma}(s,0)=\zeta_{\Gamma}(s)^{-1}$, $Z_{\Gamma}(s,1)=Z_{\Gamma}(s)=Z_{\Gamma}^{(1)}(s)$, 
$Z_{\Gamma}(s,2)=z_{\Gamma}(s-1)^{-1}=Z_{\Gamma}^{(2)}(s)$, $Z_{\Gamma}(s,r)=Z_{\Gamma}^{(r)}(s)$ and
there is a ladder structure among $\{Z_{\Gamma}(s,t)\}$ as
\begin{align}
&Z_{\Gamma}(s,t)=\frac{Z_{\Gamma}(s,t+1)}{Z_{\Gamma}(s+1,t+1)},\label{shift1}\\
&Z_{\Gamma}(s,t+1)=\prod_{n=0}^{\infty}Z_{\Gamma}(s+n,t).\nt
\end{align} 
Remark that, since $\binom{-n}{r}=(-1)^r\binom{n+r-1}{r}$ for $n>0$, 
$Z_{\Gamma}(s,-m)$ $(m\geq0)$ is written as
\begin{align*}
Z_{\Gamma}(s,-m)=&\prod_{p\in\Prim(\Gamma)}\prod_{n=0}^{m}(1-N(p)^{-s-n})^{(-1)^{n}\binom{m}{n}}
=\prod_{n=0}^{m}\zeta_{\Gamma}(s+n)^{(-1)^{n-1}\binom{m}{n}}.
\end{align*}

The main aim of this paper is to establish the analytic continuation, functional equation for $t=n\in\bZ$ 
and the determinant expressions of $Z_{\Gamma}(s,t)$ as a function of $s$ similar to \eqref{detexp} 
which allows to unify the works in, e.g. \cite{D'HP}, \cite{Sa} and \cite{KW1}. 
To describe the determinant expression, we introduce a two variable gamma function $\Gamma(s,t)$ 
via the zeta regularization which is a generalization of the multiple gamma function due to \cite{Ba} when $t\in\bZ$. 
From the multiple gamma functions, we can define multiple sine functions (see, e.g. \cite{KK})
which are important in mathematical physics; 
for instance, it has been used for expressing the solutions of the quantum Knizhnik-Zamolodchikov equation \cite{JM}.
Using these multiple sine functions together with a ``spectral determinant of sine function for $\Delta$" 
which we define again by the zeta regularized product, 
we obtain a complete higher Selberg zeta function $\hat{Z}(s,n)$, 
that is, $\hat{Z}_{\Gamma}(n-s,n)=\hat{Z}_{\Gamma}(s,n)^{(-1)^{n-1}}$ holds. 

Quite recently, Gon \cite{G2} finds an interesting relation between the period of 
a weight $4k$ holomorphic cusp form and the present higher Selberg zeta functions of several rank. 
For the Riemann zeta function case, the analysis can be done 
in the same way. See \cite{KMW} for the higher Riemann zeta function of rank 1. 

\section{Analytic continuations}
In this section, we give the analytic continuation of $Z_{\Gamma}(s,t)$ by using the Selberg trace formula 
(see, e.g. \cite{He}):
\begin{align}\label{trace}
\sum_{j\geq0}m_j\hat{f}(r_j)=&\sum_{\gamma\in \Hyp(\Gamma)}
\frac{\log{N(\delta_{\gamma})}}{N(\gamma)^{\frac{1}{2}}-N(\gamma)^{-\frac{1}{2}}}f(\log{N(\gamma)})
\nt\\+&(g-1)\int_{-\infty}^{\infty}\hat{f}(r)r\tanh{\pi r}dr.
\end{align}
where $\{r_j\}_{j\geq0}$ is the number such that 
$1/4+r_j^2=:\lambda_j$ gives the $j$-th eigenvalue of $\Delta$ on $\Gamma\bsla H$ ($\lambda_0=0$), 
$m_j$ is the multiplicity of $\lambda_j$, 
$\Hyp(\Gamma)$ is the set of the hyperbolic conjugacy classes of $\Gamma$, 
$\delta_{\gamma}$ is the element of $\Prim(\Gamma)$ such that $\gamma=\delta_{\gamma}^j$ for some $j\geq1$, 
and $f$ is a function 
whose Fourier transform $\hat{f}(r)=1/2\pi\int_{-\infty}^{\infty}f(x)e^{ixr}dx$ satisfies the condition  
$\hat{f}(r)=\hat{f}(-r)$, $\hat{f}$ is holomorphic in $\{|\Im{r}|\leq 1/2+\delta\}$, 
and $\hat{f}(r)=O(|r|^{-2-\delta})$ as $|r|\tinf$ for some $\delta>0$. 

Notice first the logarithmic derivative of $Z_{\Gamma}(s,t)$ is written as
\begin{align*}
\frac{\partial}{\partial s}\log{Z_{\Gamma}(s,t)}=\sum_{\gamma\in \Hyp(\Gamma)}
\frac{\log{N(\delta_{\gamma})}}{\big(1-N(\gamma)^{-1}\big)^{t}}N(\gamma)^{-s}.
\end{align*}
Let $f_{s,t}(x):=\big(1-e^{-|x|}\big)^{1-t}e^{(-s+1/2)|x|}$.
For $m\geq \max(2,\Re{t}+1)$, we define $f^{(m)}_{s,t}(x)$ as
\begin{align*}
f^{(m)}_{s,t}(x):=&(-1)^m\frac{\partial^m}{\partial s^m}f_{s,t}(x)
=|x|^m\big(1-e^{-|x|}\big)^{1-t}e^{(-s+1/2)|x|}.
\end{align*}
The Fourier transform of $f_{s,t}(x)$ is 
\begin{align*}
\hat{f}_{s,t}(y)=&B(2-t,s+iy-1/2)+B(2-t,s-iy-1/2),
\end{align*}
where $B(a,b)$ is the Beta function defined by $\int_{0}^{1}x^{a-1}(1-x)^{b-1}dx$.
Hence the Fourier transform of $f^{(m)}_{s,t}(x)$ is
\begin{align}
\hat{f}^{(m)}_{s,t}(y)
=&(-1)^m\frac{\partial^m}{\partial s^m}\bigg[B(2-t,s+iy-1/2)+B(2-t,s-iy-1/2)\bigg]\nt\\
=&m!\sum_{k=0}^{\infty}\binom{t+k-2}{k}\Big\{(s-1/2+k+iy)^{-(m+1)}\nt\\&+(s-1/2+k-iy)^{-(m+1)}\Big\}.\label{fourier}
\end{align}
It is easy to verify that $f_{s,t}(x)$ satisfies the required condition for the test function for the trace formula.
Hence, putting this test functions into the trace formula, we have
\begin{align}
&\frac{(-1)^m}{m!}\frac{\partial^{m+1}}{\partial s^{m+1}}\log{Z_{\Gamma}(s,t)}\nt\\
=&\sum_{j=0}^{\infty}\sum_{k=0}^{\infty}m_j\binom{t+k-2}{k}\Big\{(s-1/2+k+ir_j)^{-(m+1)}
+(s-1/2+k-ir_j)^{-(m+1)}\Big\}\nt\\
&+2(g-1)\sum_{k=0}^{\infty}\bigg\{\binom{t+k-1}{k-1}+\binom{t+k}{k}\bigg\}(s+k)^{-(m+1)}.\label{analcont}
\end{align}
The series converges absolutely and uniformly when $s$ sits apart from the obvious set of singularities.
By $m$-times integrations of the formula \eqref{analcont}, 
one shows that $\partial/\partial s\log{Z_{\Gamma}(s,t)}$ 
is analytically continued to the whole complex plane $\bC$ as a meromorphic function which has poles 
at $s=1$, $s=-k(k\geq0)$ and $s=1/2-l\pm ir_j(l\geq0,j\geq1)$ with the residues 
$1$, $\binom{t+k-1}{k+1}+\binom{t+k-2}{k}+2(g-1)\big\{\binom{t+k-1}{k-1}+\binom{t+k}{k}\big\}$ 
and $m_j\binom{t+k-2}{k}$ respectively.
Hence we obtain the following.
\begin{thm}
(i) When $t=n\in\bZ$, $Z_{\Gamma}(s,t)$ is analytically continued to the whole complex plane $\bC$ 
as a meromorphic function which has zeros (or poles) at $s=1$, 
$s=-k(k\geq0)$ and $s=1/2-l\pm ir_j(l\geq0,j\geq1)$ of order
$1$, $\binom{t+k-1}{k+1}+\binom{t+k-2}{k}+2(g-1)\big\{\binom{t+k-1}{k-1}+\binom{t+k}{k}\big\}$ 
and $m_j\binom{t+k-2}{k}$ respectively.\\
(ii) When $t\not\in\bZ$, $Z_{\Gamma}(s,t)$ is analytically continued to the whole complex plane $\bC$ 
as a function which has a simple zero at $s=1$ and branch points at $s=-k(k\geq0)$, $1/2-l\pm ir_j(l\geq0,j\geq1)$, 
and is non-zero holomorphic at any other points.
\end{thm}

\section{Determinant expressions} 
The aim of this section is to generalize the determinant expression \eqref{detexp} of $Z_{\Gamma}(s)$. 
First, in order to determine the gamma factor, we prove the following lemma.
\begin{lem}\label{lem1}
Let
\begin{align*}
\zeta_t(z,s):=&\sum_{k=0}^{\infty}\binom{t+k-1}{k}(k+s)^{-z}.
\end{align*}
Then the functions $\zeta_t(z,s)$ is 
analytically continued to the whole complex plane $\bC$ as meromorphic function of $z$
which has only simple pole at $z=t$.
\end{lem}

\begin{proof}
Put
\begin{align*}
\theta_t(x,s):=&\sum_{k=0}^{\infty}\binom{t+k-1}{k}e^{-(k+s)x}=\frac{e^{-sx}}{(1-e^{-x})^{t}}.
\end{align*}
It is easy to see that
\begin{align}
\zeta_t(z,s)=&\frac{1}{\Gamma(z)}\int_{0}^{\infty}\theta_t(x,s)x^{z-1}dx\nt\\
=&\frac{\Gamma(1-z)e^{\pi iz}}{2\pi i}\int_{C}\theta_t(x,s)x^{z-1}dx, \label{zett}
\end{align}
where $\Gamma(z):=\int_{0}^{\infty}e^{-t}t^{z-1}dt$ and
\begin{align*}
C:=&\{x;\infty\to\delta\}\cup\{\delta e^{i\theta}|\theta;0\to2\pi i\}\cup\{x;\delta e^{2\pi i}\to\infty e^{2\pi i}\}.
\end{align*} 
The expression \eqref{zett} gives immediately the analytic continuation of $\zeta_t(z,s)$.
\end{proof}

Since $\zeta_{t}(z,s)$ is holomorphic at $z=0$, we can define $\Gamma(s,t)$as follows. 
\begin{definition}
\begin{align}
\Gamma(s,t):=&\exp\Big(\frac{\partial}{\partial z}\zeta_t(z,s)\Big|_{z=0}\Big).
\end{align}
\end{definition}
The function $\Gamma(s,t)$ is shown to be meromorphic with respect to $s\in\bC$ 
when $t\in\bZ$ in the same way as \cite{KiW} 
and is one of the generalization of the multiple Gamma functions; in fact, 
\begin{align*}
\Gamma(s,0)=1/s, \quad \Gamma(s,1)=\Gamma(s)/\sqrt{2\pi}
\end{align*}
by the Lerch formula \cite{Le}
and, for $t=n\geq2$, $\Gamma(s,n)$ essentially coincides with 
the multiple Gamma functions (see, e.g. \cite{Ba}). 
When $t=n\in\bZ_{\geq0}$, the following multiple sine functions are defined in terms of $\Gamma(s,n)$ (see \cite{KK}).
\begin{align*}
\cS(s,n):=&\Gamma(s,n)^{-1}\Gamma(n-s,n)^{(-1)^n}.
\end{align*}
It is easy to see from the definition that the function $\Gamma(s,t)$ satisfies  
\begin{align}
&\frac{\Gamma(s,t+1)}{\Gamma(s+1,t+1)}=\Gamma(s,t)\label{shift2}
\end{align}
and 
\begin{align}
\frac{(-1)^m}{m!}&\frac{\partial^{m+1}}{\partial s^{m+1}}\log{\Gamma(s,t)}
=-\sum_{k=0}^{\infty}\binom{t+k-1}{k}(s+k)^{-(m+1)}.
\end{align}
Hence, the formula \eqref{analcont} is expressed as 
\begin{align}
&\frac{\partial^{m+1}}{\partial s^{m+1}}\log{Z_{\Gamma}(s,t)}\nt\\
=&-2(g-1)\frac{\partial^{m+1}}{\partial s^{m+1}}\log{\Big(\Gamma(s+1,t+1)\Gamma(s,t+1)\Big)}\nt\\
&-\sum_{j=0}^{\infty}\frac{\partial^{m+1}}{\partial s^{m+1}}
\log\Big(\Gamma(s-1/2+ir_j,t-1)\Gamma(s-1/2-ir_j,t-1)\Big).\label{det0}
\end{align}
Therefore, what we expect here is to give an expression of $Z_{\Gamma}(s,t)$ as in the form
\begin{align}
``Z_{\Gamma}(s,t)=&\exp{P_{\Gamma}(s,t)}\Big(\Gamma(s+1,t+1)\Gamma(s,t+1)\Big)^{-2(g-1)}\nt\\
&\times\Big(\prod_{j=0}^{\infty}\Gamma(s-1/2+ir_j,t-1)\Gamma(s-1/2-ir_j,t-1)\Big)^{-1}",\label{det1}
\end{align}
where $P_{\Gamma}(s,t)$ is a polynomial whose degree is less than or equals to $\max{(2,\Re{t}+1)}$.
The product over $r_j$ however does not converge.
Hence, we need to regularize the spectral factor of \eqref{det1} as a certain suitable product. 

Now, we recall the following elementary fact:
``the determinant of the finite dimensional linear operator $A$ coincides the product of the eigenvalues of $A$".
In this sense, although $\Delta$ is not finite dimensional, 
the spectral factor of \eqref{det1} can be formally interpreted as 
\begin{align}
&``\det\Big(\Gamma(s-1/2+\sqrt{1/4-\Delta},t-1)\Gamma(s-1/2-\sqrt{1/4-\Delta},t-1)\Big)".\label{deter}
\end{align}
When $t=1$, the determinant \eqref{deter} has been justified as follows (see \cite{Sa}).
\begin{align}
\det(\Delta+s(1-s)):=\exp{\Big(-\frac{\partial}{\partial z}\zeta_{\Delta}(z,s(1-s))\Big|_{z=0}\Big)},\label{det}
\end{align}
where 
\begin{align*}
\zeta_{\Delta}(z,x):=\sum_{j=1}^{\infty}(\lambda_j+x)^{-z}\quad\Re{s}>1.
\end{align*}
Since the function $\zeta_{\Delta}(z,x)$ is analytically continued to the whole complex plane $\bC$ 
which is holomorphic at $s=0$, \eqref{det} is actually a ``well-defined" one . 
In order to give the generalization of \eqref{det}, we prove the following lemma.
\begin{lem}\label{lem2}
Let 
\begin{align*}
\xi_{\Gamma,t}^{(+)}(z,s)
:=&\sum_{j=0}^{\infty}\sum_{k=0}^{\infty}\binom{t+k-1}{k}(s-\frac{1}{2}+k+ir_j)^{-z},\\
\xi_{\Gamma,t}^{(-)}(z,s)
:=&\sum_{j=0}^{\infty}\sum_{k=0}^{\infty}\binom{t+k-1}{k}(s-\frac{1}{2}+k-ir_j)^{-z}.
\end{align*}
The functions $\xi_{\Gamma,t}^{(+)}(z,s)$ and $\xi_{\Gamma,t}^{(-)}(z,s)$ are 
analytically continued to the whole complex plane $\bC$ as meromorphic functions of $z$
which have only simple poles at $z=t+2$.
\end{lem}

\begin{proof}
Put
\begin{align*}
\phi_{\Gamma,t}^{(+)}(x,s)
:=&\sum_{j=0}^{\infty}\sum_{k=0}^{\infty}\binom{t+k-1}{k}e^{-x(s'-ie^{i\theta}k+e^{i\theta}r_j)}\\
=&e^{-xs}(1-e^{ie^{i\theta}x})^{-t}\sum_{j=0}^{\infty}e^{-e^{i\theta}r_j x}
\end{align*}
for some $\theta (0<\theta<\pi/2)$ and $s'=-ie^{i\theta}(s-1/2)$.
It is easy to see that 
\begin{align}
\xi_{\Gamma,t}^{(+)}(z,s)
=&\frac{e^{iz(\theta-\pi/2)}}{\Gamma(s)}\int_{0}^{\infty}\phi_{\Gamma,t}^{(+)}(x,s)x^{z-1}dx\nt\\
=&\frac{e^{iz(\theta-\pi/2)}\Gamma(1-s)}{2\pi i}\int_{C}\phi_{\Gamma,t}^{(+)}(x,s)x^{z-1}dx.\label{xi}
\end{align}
Applying the method used in \cite{CV}, we can see that the function
\begin{align*}
\sum_{j=0}^{\infty}e^{-e^{i\theta}r_j x}
\end{align*}
has a double pole at $x=0$. 
Hence, the analytic continuation of $\xi_{\Gamma,t}^{(+)}(z,s)$ follows from \eqref{xi}.
Similarly, the analytic continuation of $\xi_{\Gamma,t}^{(-)}(z,s)$ can be obtained.
\end{proof}

Hence, the following regularized determinant can be defined.
\begin{definition}
Let $t\in\bC$ and $n\in\bZ$.
We define $\det\Gamma$ and $\det\cS$ as follows.
\begin{align*}
&\det\Gamma(s-1/2\pm\sqrt{1/4-\Delta},t):=\exp{\Big(\Res_{z=0}\xi_{\Gamma,t}^{(\pm)}(z,s)/z^2\Big)},\\
&\det\cS(s-1/2\pm\sqrt{1/4-\Delta},n)\\:=&\det\Gamma(s-1/2\pm\sqrt{1/4-\Delta},n)^{-1}
\det\Gamma(n-s-1/2\mp\sqrt{1/4-\Delta},n)^{(-1)^n}.
\end{align*}
\end{definition}
Obviously, when $\xi_{\Gamma,t}^{(\pm)}(z,s)$ is holomorphic at $z=0$, we have
\begin{align*}
\Res_{z=0}\xi_{\Gamma,t}^{(\pm)}(z,s)/z^2=\frac{\partial}{\partial z}\xi_{\Gamma,t}^{(\pm)}(z,s)\Big|_{z=0}.
\end{align*}
The same discussion developed in \cite{KW1} (see also \cite{KiW}) shows that 
these determinants satisfy that
\begin{align}
&\frac{\det{\Gamma(s-1/2\pm\sqrt{1/4-\Delta},t+1)}}{\det{\Gamma(s+1/2\pm\sqrt{1/4-\Delta},t+1)}}
=\det{\Gamma(s-1/2+\sqrt{1/4-\Delta},t)},\label{shift3}\\
&\frac{(-1)^m}{m!}\frac{\partial^{m+1}}{\partial s^{m+1}}\log{\det{\Gamma(s-1/2\pm\sqrt{1/4-\Delta},t)}}\nt\\
=&-\sum_{j=0}^{\infty}\sum_{k=0}^{\infty}\binom{t+k-1}{k}(s-1/2\pm ir_j)^{-(m+1)}.
\end{align}
Hence, it follows that \eqref{analcont} is expressed as 
\begin{align}
\frac{\partial^{m+1}}{\partial s^{m+1}}\log{Z_{\Gamma}(s,t)}
=&-2(g-1)\frac{\partial^{m+1}}{\partial s^{m+1}}\log{\Big(\Gamma(s+1,t+1)\Gamma(s,t+1)\Big)}\nt\\
&-\sum_{j=0}^{\infty}\frac{\partial^{m+1}}{\partial s^{m+1}}
\log\Big(\det\Gamma(s-1/2+\sqrt{1/4-\Delta},t-1)\nt\\
&\times\det\Gamma(s-1/2-\sqrt{1/4-\Delta},t-1)\Big).\label{det01}
\end{align}
Therefore, we have
\begin{align}
Z_{\Gamma}(s,t)=&\exp{\big(P_\Gamma(s,t)\big)}\Big(\Gamma(s+1,t+1)\Gamma(s,t+1)\Big)^{-2(g-1)}\nt\\
&\times\Big(\det\Gamma(s-1/2+\sqrt{1/4-\Delta},t-1)\nt\\
&\times\det\Gamma(s-1/2-\sqrt{1/4-\Delta},t-1)\Big)^{-1},\label{det2}
\end{align}
where $P_{\Gamma}(s,t)$ is a of degree $\leq\max{(2,\Re{t}+1)}$. 
One may assume that $P_{\Gamma}(s,t)\in\bR$ for $s,t\in\bR$.
Let 
\begin{align*}
P_{\Gamma}(s,t)=\sum_{l=0}^{\max{(2,[\Re{t}+1])}}p_l^{(t)}s^l.
\end{align*}
By virtue of \eqref{shift1}, \eqref{shift2} and \eqref{shift3}, we have
\begin{align}
P_{\Gamma}(s,t+1)-P_{\Gamma}(s+1,t+1)=P_{\Gamma}(s,t).\label{shift4}
\end{align}
Hence, it is easy to get 
\begin{align*}
P_{\Gamma}(s,t)=\begin{cases}\disp\sum_{l=0}^{[\Re{t}+1]}p_l^{(t)}s^l&\text{for $t\geq-1$},\\
0&\text{for $t<-1$}.\end{cases}
\end{align*} 
Comparing the coefficients of the both sides of \eqref{shift4}, we obtain 
\begin{align*}
p_l^{(t)}=-\sum_{m=l+1}^{[\Re{t}+1]}\binom{m}{l}p_m^{(t+1)}\quad\text{for $0\leq l\leq \Re{t}+1$}.\label{coef}
\end{align*}
Hence we have
\begin{align}
p_{l+1}^{(t+1)}=&\sum_{m=l}^{[\Re{t}+1]}(-1)^{l+m+1}\binom{m}{l}p_l^{(t)}\quad\text{for $0\leq l\leq \Re{t}+1$}.
\end{align}
Therefore, we obtain the following theorem.
\begin{thm}\label{detthm}
The function $Z_{\Gamma}(s,t)$ is expressed as 
\begin{align}
Z_{\Gamma}(s,t)
&=\exp{\big(P_\Gamma(s,t)\big)}\Big(\Gamma(s+1,t+1)\Gamma(s,t+1)\Big)^{-2(g-1)}\nt\\
&\times\Big(\det\Gamma(s-1/2+\sqrt{1/4-\Delta},t-1)\nt\\&\times\det\Gamma(s-1/2-\sqrt{1/4-\Delta},t-1)\Big)^{-1},
\end{align}
where $P_{\Gamma}(s,t)$ is a polynomial of degree $\leq \Re{t}+1$ when $\Re{t}\geq-1$ and equals to $0$ when $\Re{t}<-1$. 
When $\Re{t}\geq -1$, the coefficients of $P_{\Gamma}(s,t)=\sum_{l=0}^{[\Re{t}+1]}p_l^{(t)}s^l$ is determined 
recursively as follows.
\begin{align*}
p_0^{(t)}=&\lim_{s\to0}\bigg[\log{Z_{\Gamma}(s,t)}+\frac{\partial}{\partial z}
\Big\{2(g-1)\zeta_t(z,s)+\xi_{\Gamma,t}^{(+)}(z,s)+\xi_{\Gamma,t}^{(-)}(z,s)\Big\}\Big|_{z=0}\bigg],\\
p_{l+1}^{(t+1)}=&\sum_{m=l}^{[\Re{t}+1]}(-1)^{l+m+1}\binom{m}{l}p_l^{(t)}\quad\text{for $0\leq l\leq \Re{t}+1$}.
\end{align*}
\end{thm}

\section{Functional equations}
We establish here the functional equation of $Z_{\Gamma}(s,t)$ for $t\in\bZ$.
First, we prove the following lemma.
\begin{lem}\label{lemfe}
Let $n$ be an integer. Then we have
\begin{align}
&\hat{f}_{n-s,n}(y)=(-1)^n\hat{f}_{s,n}(y)&\text{for $n\leq1$},\label{test1}\\
&\hat{f}_{n-s,n}(y)+\frac{(-1)^n\pi}{(n-2)!}\tan{\pi(n-s+iy)}\prod_{k=1}^{n-2}(n-s+iy-k+1/2)\nt\\
&=(-1)^n\hat{f}_{s,n}(y)+\frac{\pi}{(n-2)!}\tan{\pi(s+iy)}
\prod_{k=1}^{n-2}(s+iy-k+1/2)&\text{for $n\geq2$}.\label{test2}
\end{align}
Here $\hat{f}_{s,n}(y)$ is given in Section 2.
\end{lem}

\begin{proof}
Since 
\begin{align*}
\hat{f}_{s,t}(y)=&B(2-t,s+iy-1/2)+B(2-t,s-iy-1/2),
\end{align*}
we have
\begin{align*}
\hat{f}_{t-s,t}(y)=&B(2-t,t-s+iy-1/2)+B(2-t,t-s-iy-1/2)\\
=&B(2-t,s+iy-1/2)\frac{\sin{\pi(s+iy-1/2)}}{\sin{\pi(s+iy-1/2-t)}}\\
&+B(2-t,s-iy-1/2)\frac{\sin{\pi(s-iy-1/2)}}{\sin{\pi(s-iy-1/2-t)}}.
\end{align*}
Using the addition theorem of the sine function, we obtain
\begin{align}
\hat{f}_{t-s,t}(y)=&\cos{(\pi t)}\hat{f}_{s,t}(y)-\sin{(\pi t)}\big\{\tan{\pi(s+iy-t)}B(2-t,s+iy-1/2)\nt\\
&+\tan{\pi(s-iy-t)}B(2-t,s-iy-1/2)\big\}\nt\\
=&\cos{(\pi t)}\hat{f}_{s,t}(y)+\frac{\pi}{\Gamma(t-1)}
\bigg\{\tan{\pi(s+iy-1/2)}\frac{\Gamma(s+iy-1/2)}{\Gamma(s+iy+3/2-t)}\nt\\
&+\tan{\pi(s-iy-1/2)}\frac{\Gamma(s-iy-1/2)}{\Gamma(s-iy+3/2-t)}\bigg\}.
\end{align}
Hence, the desired result follows.  
\end{proof}

Put 
\begin{align}
&H_{\Gamma}(s,t):=Z_{\Gamma}(s,t)\Big(\Gamma(s+1,t+1)\Gamma(s,t+1)\Big)^{2(g-1)}.
\end{align}
According to \eqref{analcont}, we have
\begin{align}
\frac{(-1)^m}{m!}\frac{\partial^{m+1}}{\partial s^{m+1}}\log{H_{\Gamma}(s,t)}
=&\sum_{j=0}^{\infty}\frac{1}{m!}\hat{f}_{s,t}^{(m)}(r_j)
=\frac{(-1)^m}{m!}\sum_{j=0}^{\infty}\frac{\partial^{m}}{\partial s^m}\hat{f}_{s,t}(r_j).\label{analcont2}
\end{align}
Because of \eqref{analcont2} and Lemma \ref{lemfe}, the following formulas hold.
\begin{align}
&(-1)^m\frac{\partial^{m+1}}{\partial s^{m+1}}\log{H_{\Gamma}(n-s,n)}
=(-1)^n\frac{\partial^{m+1}}{\partial s^{m+1}}\log{H_{\Gamma}(s,n)}&\text{for $n\leq1$},\label{functeq1}
\end{align}
\begin{align}
&(-1)^m\bigg[\frac{\partial^{m+1}}{\partial s^{m+1}}\log{H_{\Gamma}(n-s,n)}\nt\\
&+\frac{(-1)^n\pi}{(n-2)!}\sum_{j=0}^{\infty}\frac{\partial^{m}}{\partial s^{m}}
\Big\{\tan{\pi(n-s+ir_j)}\prod_{k=1}^{n-2}(n-s+ir_j-k-1/2)\Big\}\bigg]\nt\\
=&(-1)^n\frac{\partial^{m+1}}{\partial s^{m+1}}\log{H_{\Gamma}(s,n)}\nt\\
&+\frac{\pi}{(n-2)!}\sum_{j=0}^{\infty}\frac{\partial^{m}}{\partial s^{m}}
\Big\{\tan{\pi(s+ir_j)}\prod_{k=1}^{n-2}(s+ir_j-k-1/2)\Big\}\quad\text{for $n\geq2$}.\label{functeq2}
\end{align}
Let
\begin{align*}
\tilde{H}_{\Gamma}(s,n):=&H(s,n)\det\Gamma(s-1/2-\sqrt{1/4-\Delta},n-1)\nt\\
&\times\det\Gamma(n-s-1/2+\sqrt{1/4-\Delta},n-1)^{(-1)^n}.
\end{align*}
Then we can express \eqref{functeq2} as
\begin{align}
&(-1)^m\frac{\partial^{m+1}}{\partial s^{m+1}}\log{\tilde{H}_{\Gamma}(n-s,n)}
=(-1)^n\frac{\partial^{m+1}}{\partial s^{m+1}}\log{\tilde{H}_{\Gamma}(s,n)}\quad\text{for $n\geq2$}.\label{functeq3}
\end{align}
Hence \eqref{functeq1} and \eqref{functeq3} yield
\begin{align*}
H_{\Gamma}(n-s,n)=H_{\Gamma}(s,n)^{(-1)^{n-1}}
&\times\exp{\text{(polynomial)}}& \text{for $n\leq1$},\\
\tilde{H}_{\Gamma}(n-s,n)=\tilde{H}_{\Gamma}(s,n)^{(-1)^{n-1}}
&\times\exp{\text{(polynomial)}}& \text{for $n\geq2$}.
\end{align*}
On account of Theorem \ref{detthm}, we obtain the following functional equations.
\begin{thm}
Let $n$ be an integer. Then we have
\begin{align*}
H_{\Gamma}(n-s,n)\exp{P_{\Gamma}(n-s,n)}
&=\Big(H_{\Gamma}(s,n)\exp{P_{\Gamma}(s,n)}\Big)^{(-1)^{n-1}}& \text{for $n\leq1$},\\
\tilde{H}_{\Gamma}(n-s,n)\exp{P_{\Gamma}(n-s,n)}
&=\Big(\tilde{H}_{\Gamma}(s,n)\exp{P_{\Gamma}(s,n)}\Big)^{(-1)^{n-1}}& \text{for $n\geq2$},
\end{align*}
where $P_{\Gamma}(s,t)$ is given in Theorem \ref{detthm}.
\end{thm}
The theorem above allows to give the definition of a complete higher Selberg zeta function 
by use of multiple sine functions.
\begin{cor}
Let $n$ be an integer. We define the complete higher Selberg zeta function $\hat{Z}_{\Gamma}(s,n)$ as follows.
\begin{align*}
\hat{Z}_{\Gamma}(s,n):=&Z_{\Gamma}(s,n)\cS(s,n+1)^{-2(g-1)}\exp{P_{\Gamma}(s,n)}&\text{for $n\leq1$},\\
\hat{Z}_{\Gamma}(s,n):=&Z_{\Gamma}(s,n)\cS(s,n+1)^{-2(g-1)}\exp{P_{\Gamma}(s,n)}\\ 
&\times\det\cS(s-1/2-\sqrt{1/4-\Delta},n-1)^{-1}&\text{for $n\geq2$}.
\end{align*}
Then we have
\begin{align*}
\hat{Z}_{\Gamma}(n-s,n)=&\hat{Z}_{\Gamma}(s,n)^{(-1)^{n-1}}.
\end{align*}
\end{cor}

The theorem and corollary above give, in particular, an affirmative answer of the question raised at p.465 in \cite{KW2} 
for the Selberg zeta function.

\noindent 
\text{HASHIMOTO, Yasufumi}\\ 
Graduate School of Mathematics, Kyushu University.\\  6-10-1, Hakozaki, Fukuoka, 812-8581 JAPAN.\\ 
\text{hasimoto@math.kyushu-u.ac.jp}\\ 

\noindent 
\text{WAKAYAMA, Masato}\\ 
Faculty of Mathematics, Kyushu University.\\  
6-10-1, Hakozaki, Fukuoka, 812-8581 JAPAN.\\ 
\text{wakayama@math.kyushu-u.ac.jp}\\

\end{document}